\documentclass[aps,prb,preprint,superscriptaddress,amsmath,amssymb,onecolumn]{revtex4-2}

\usepackage{graphicx}
\usepackage{dcolumn}
\usepackage{bm}
\usepackage[colorlinks,linkcolor=blue,anchorcolor=blue,citecolor=blue] {hyperref}
\usepackage{hyperref}
\usepackage{multirow}
\usepackage{lineno}

\begin{document}

\title{A General Method to Design Acoustic Higher-Order Topological Insulators}

\author{An-Yang Guan}
\thanks{These authors have contributed equally to this work.}
\author{Zhang-Zhao Yang}
\thanks{These authors have contributed equally to this work.}
\affiliation{Key Laboratory of Modern Acoustics, MOE, Institute of Acoustics, Department of Physics, Collaborative Innovation Center of Advanced Microstructures, Nanjing University, Nanjing 210093, People’s Republic of China}
\author{Xin-Ye Zou}
\email{xyzou@nju.edu.cn}
\author{Jian-Chun Cheng}
\affiliation{Key Laboratory of Modern Acoustics, MOE, Institute of Acoustics, Department of Physics, Collaborative Innovation Center of Advanced Microstructures, Nanjing University, Nanjing 210093, People’s Republic of China}
\affiliation{State Key Laboratory of Acoustics, Chinese Academy of Sciences, Beijing 100190, People’s Republic of China}

\begin{abstract}
Acoustic systems that are without limitations imposed by the Fermi level have been demonstrated as significant platform for the exploration of fruitful topological phases. By surrounding the nontrivial domain with trivial “environment”, the domain-wall topological states have been theoretically and experimentally demonstrated. In this work, based on the topological crystalline insulator with a kagome lattice, we rigorously derive the corresponding Hamiltonian from the traditional acoustics perspective, and exactly reveal the correspondences of the hopping and onsite terms within acoustic systems. Crucially, these results directly indicate that instead of applying the trivial domain, the soft boundary condition precisely corresponds to the theoretical models which always require generalized chiral symmetry. These results provide a general platform to construct desired acoustic topological devices hosting desired topological phenomena for versatile applications. 
\end{abstract}

\maketitle

\section{\label{sec_1}introduction}
Originating from the condensed matter physics, the concept of topological insulators (TIs)
\cite{cm_1,cm_2,cm_3} have been intensely investigated in the classical wave systems \cite{cw_1,cw_2,cw_3,cw_4,cw_5,cw_6,cw_7,cw_8,cw_9,cw_10,cw_11,cw_12,cw_13,cw_14,Deng1,cw_15,cw_16,cw_17,cw_18}. More recently, as a counterpart of topological insulators in solid materials, the topological crystalline insulators (TCIs) that can host higher-order topological states have attracted growing attention \cite{tci_2,tci_3,Slager1,Slager2,tci_1}. Distinct from the conventional TIs that are induced by external fields \cite{ef_1,ef_2,ef_3} or strong spin-orbit couplings \cite{tis_1,tis_2,tis_3,tis_4,tis_5,tis_6,tis_7}, the topological invariants of the TCIs are determined by the symmetry of the lattice, and the emerging topological states are protected by certain lattice symmetries. Importantly, some reported works have theoretically and experimentally demonstrated that the shapes of the TCIs can also impact the generation of the topological states, even if that the bulk of the structure is nontrivial \cite{YZZ,ba_1,ba_2,ba_3,ba_4}.

Generally, in quantum systems, the vacuum can be regarded as a trivial domain surrounding the open boundaries of the nontrivial structure \cite{cm_1}. However, for the analogous TCIs in classical wave systems, i.e., acoustic systems, it is the boundary conditions such as hard boundary or soft boundary that we imposed to construct finite structures, with which the natural behaviors of the emerging states on the boundaries are distinct \cite{add1}. On the other hand, different from the existing theoretical models focusing on studying the `zero-energy’ modes, in real physical systems, the diagonal terms of the corresponding Hamiltonian, which represent the on-site energy of the `virtual’ atoms, are always non-negligible. Due to particle-hole symmetry, such the terms in the bulk lattices are always identical; however, the reduced symmetry on the boundaries of the finite structure can lead to extra hopping on the boundary lattices, which directly impacts the corresponding on-site energy. Therefore, how does the on-site energy affect the Hamiltonian, namely, what actually are the impacts of the boundaries on the generation of the topological states in classical wave systems, is still an open question. 

The purpose of this work is to propose a general platform to realize the analogy of the TCIs in acoustic systems. From the acoustic-electric analogy approach, we stringently deduce the Hamiltonian of the resonant system with a two-dimensional kagome lattice from the perspective of traditional acoustics, so as to unveil the connection between the intra- or inter-cell hoppings in the theoretical model and the acoustic parameters in real physical systems. Crucially, by presenting a finite time-reversal-invariant topological structure, we demonstrate that different boundary conditions correspond to distinct on-site energy in edge lattices. In addition to the nontrivial topological phase, the existence of the topological states requires that the boundaries of the system shall be soft. All the theoretical predictions are precisely supported by numerical simulations using the finite-element method. These results not only reveal the rigorous analogy of the TIs from condensed matter physics to acoustic systems, but also provide a general platform for the design of acoustic topological systems with topological states at any desired frequencies.

\section{\label{sec_2}Topological phase of the bulk bands in a kagome lattice}
\begin{figure}[htbp]
\centering
\includegraphics[width=15cm]{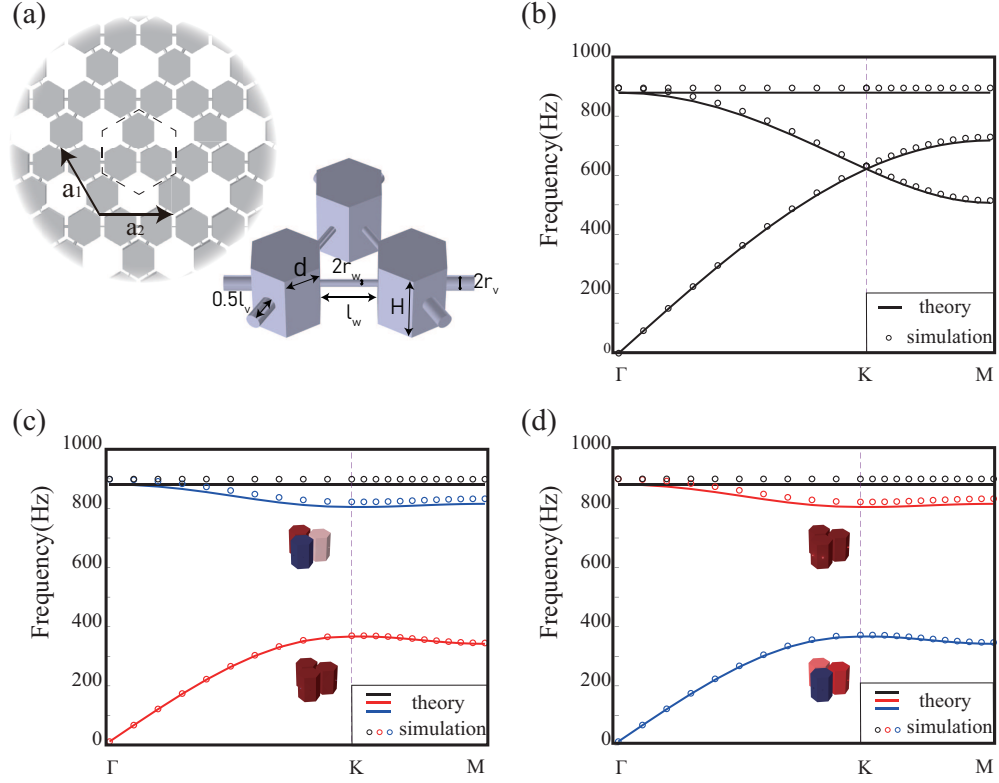}
\caption{Topological phase transition of infinite periodic kagome structure.
(a) Schematic of acoustic kagome lattice composed of cavities and tubes.
(b) Energy band structure when it comes to the critical point of topological transition when $v/w = 1$. (c) Trivial energy band structure when $v/w = 0.2$. (d) Nontrivial energy band structure when $v/w = 5$. The hollow dots are the simulation results and the lines are the theoretical results. }
\label{fig_1}
\end{figure}

We start from a kagome lattice that consists of three identical hexagonal Helmholtz resonators (labeled with 1-3, respectively) connected by waveguide tubes as shown in Fig. \hyperref[fig_1]{1(a)}, and consider that due to $C_3$ symmetry and translation invariant, all the resonators are identical. The side length and the height of the resonators are $d = 10 {\rm mm}$ and $ h = 25{\rm mm}$, respectively. The length of the tubes is $ l_{w} = l_{v} = 2.5{\rm mm}$,and $V$ is the volume of the cavity. The mass density of the air and the corresponding sound speed are  defined as $\rho =  1.23{\rm kg/m^3}$ and $ c =  343{\rm m/s}$, respectively. Generally, the bands are predicted to invert along with the change of inter-cell hoppings and intra-cell hoppings \cite{tpk_1,tpk_2,tpk_3}, which in this model are determined by the radius of the inter-cell tubes $r_w$ and intra-cell tubes $r_v$, respectively.

Figures \hyperref[fig_1]{1(b)-1(d)} depict the energy band structures of the lattice when $r_w = r_v = 0.55{\rm mm}$, $r_w=0.75{\rm mm}$, $r_v = 0.3{\rm mm}$ and $r_w = 0.3{\rm mm}$, $r_v=0.75{\rm mm}$, respectively. It is obvious to see that a symmetry-protected Dirac cone exists when $r_w = r_v$, which indicates a critical point of topological phase transition. For $ C_3 $-symmetric lattice, the bulk polarization $ p_i $ is defined as \cite{cw_8}
\begin{equation}
e^{-i \pi p_i}=\prod _{n\in \mathrm{occ}}\frac{\theta_n(\mathbf K)}{\theta_n(\mathbf \Gamma )}
\label{eq_1}
\end{equation}
where $\theta_n(\mathbf k) = \left \langle u_n(\mathbf k)|R_3|u_n(\mathbf k)\right \rangle$is calculated by the three-fold symmetric operator $R_3$ (rotation by $2\pi /3$) applied to the corresponding Bloch wave function $u_n(\mathbf k)$ at high-symmetry points of the lattice. Since that there is only one band below the band gap,the bulk polarization can be obtained as
\begin{equation}
\left ( p_1,p_2 \right )=\left\{\begin{matrix}
\left (  -1/3,-1/3\right ),w<v\\ \left (0,0  \right ),w>v
\end{matrix}\right.
\label{eq_2}
\end{equation}
As a result, the non-zero polarization when $w<v$ indicates the nontrivial topological phase of the band, and the case $w<v$ indicates the trivial phase. Since that all the atoms are considered identical, the boundary-induced filling anomaly is predicted to be induced along with the nontrivial bulk polarization case, which is represented by the existence of the topological edge states when the boundaries are open. Meanwhile, the corner-induced filling anomaly can also be induced that is characterized by the lower-dimensional topological corner states \cite{cs_1}.

We have discussed the case when all the atoms are identical in the thermodynamic limit. However, in the next section, we show that when it comes to closed classical systems, the on-site energy of the system is always affected by the boundaries.

\section{\label{sec_3}Diagonal terms of Hamiltonian and boundary correspondences}
\begin{figure}[htbp]
\centering
\includegraphics[width=15cm]{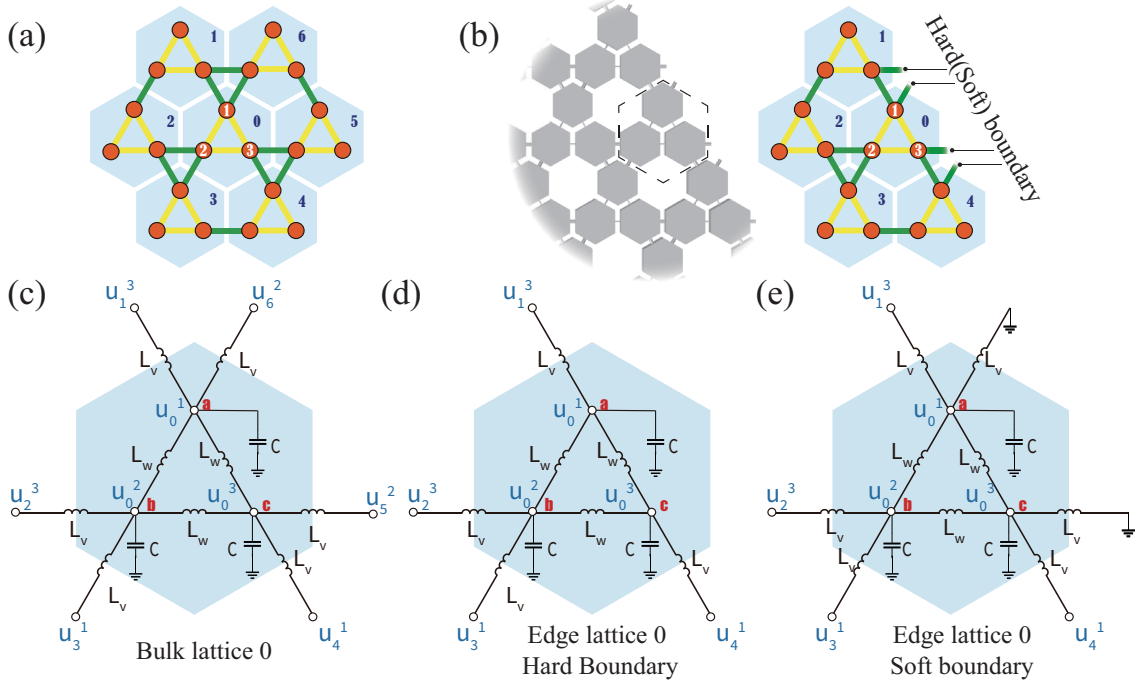}
\caption{ (a) Schematic of the bulk lattice with six nearest-neighbors lattices. (b) Schematic of the edge lattice with four nearest-neighbor lattices. The outmost tubes are connected to hard (soft) boundaries. (c) The lumped circuit model of the bulk lattice 0 in (a). (d) The lumped circuit model of the edge lattice 0 in (b) with hard boundary. (e) The lumped circuit model of the edge lattice 0 in (b) with soft boundary.}
\label{fig_2}
\end{figure}

Here, we stringently derive the precise expression of Hamiltonian of real acoustic systems, which shows that the effects of different boundary conditions are reflected in diagonal terms of Hamiltonian. For the convenience of elaboration, the bulk lattice with labeled cavities and its six nearest neighbors are depicted in Fig. \hyperref[fig_2]{2(a)}, when the corresponding circuit counterpart is presented in Fig. \hyperref[fig_2]{2(c)}. The cavities serve as capacitors $C = V /(\rho c^2)$ and the inter-cell (intra-cell) tubes act as inductors $L_{v(w)} = \rho (l_{v(w)}+1.7r_{v(w)} )/ (\pi r_{v(w)}^2)$.\cite{analogy} Based on Kirchhoff's current law at the three points(respectivly labeled with a,b and c in Fig. \hyperref[fig_2]{2(c)}), we obtain
\begin{subequations}
\begin{equation}
-\left ( 2w+2v \right ) u_{0}^{1} + w u_{0}^{2} + w u_{0}^{3} + v u_{6}^{2} + v u_{1}^{3} = \omega^2 u_{0}^{1}
\end{equation}
\begin{equation}
-\left ( 2w+2v \right ) u_{0}^{2} + w u_{0}^{3} + w u_{0}^{1} + v u_{2}^{3} + v u_{3}^{1} = \omega^2 u_{0}^{2}
\end{equation}
\begin{equation}
-\left ( 2w+2v \right ) u_{0}^{3} + w u_{0}^{1} + w u_{0}^{2} + v u_{4}^{1} + v u_{5}^{2} = \omega^2 u_{0}^{3}
\end{equation}
\label{eq_3}
\end{subequations}
where $u_m^n$ represents the sound pressure in the n-th cavities of m-th lattice and $v = -1/L_vC$, $w = -1/L_w C$. For a periodic structure which guarantees that $u_m^n$ can be described as Bloch wave function, Eq.\hyperref[eq_3]{(3)} can be rewritten as the vector form 
\begin{equation}
\mathcal H_{0} \mathbf u = \omega^2 \mathbf u
\label{eq_4}
\end{equation}
where $\mathbf u = \left[ u_{0}^{1}\ u_{0}^{2}\ u_{0}^{3}\right]^{\rm T}$, and the $\mathcal H_{0}$ then can be given as
\begin{equation}
\mathcal H_{0}(\mathbf k) =
\left[
\begin{array}{ccc}
    -2w-2v & w+ve^{j\mathbf k\cdot (\mathbf a_1+ \mathbf a_2)} & w+ve^{j\mathbf k \cdot \mathbf a_1} \\
    w+ve^{-j\mathbf k\cdot (\mathbf a_1+ \mathbf a_2)} & -2w-2v & w+ve^{-j\mathbf k \cdot \mathbf a_2} \\
    w+ve^{-j\mathbf k \cdot \mathbf a_1} & w+ve^{j\mathbf k \cdot \mathbf a_2} & -2w-2v
\end{array}
\right]
\label{eq_5}
\end{equation}
where $\mathbf k$ is the Bloch wave vector and $\mathbf a_1,\mathbf a_2$ represent lattice constant. It is obvious to see that $\mathcal H_{0}$ is the direct Hamiltonian of the lattice. Therefore,by solving Eq.\hyperref[eq_3]{(3)}, the band structure can be obtained as shown in Figs. \hyperref[fig_1]{1(b)-1(d)}.

We note that for the periodic systems, the diagonal terms of the corresponding Hamiltonian of the structure, which can be derived from Eq.\hyperref[eq_4]{(4)}, are identical. Whereas, if the boundaries of the systems are considered, the difference of diagonal terms in Hamiltonian of the whole structure do have a tremendous effect on the topological properties. As depicted in Fig. \hyperref[fig_2]{2(b)}, we assume that there is a lattice 0 located in the edge of a finite structure. Compared to Fig. \hyperref[fig_2]{2(a)}, the neighbor lattice 5 and 6 in Fig. \hyperref[fig_2]{2(a)} are replaced by the hard or soft boundary that the outmost tubes connected to. In the lumped circuit model, the hard boundary is equivalent to disconnection case while the soft boundary is equivalent to being grounded, illustrated in Figs. \hyperref[fig_2]{2(d)} and \hyperref[fig_2]{2(e)} respectively. Accordingly,for the hard boundary case, the equations of edge lattice can be obtained as:
\begin{subequations}
\begin{equation}
-\left ( 2w+v \right ) u_{0}^{1} + w u_{0}^{2} + w u_{0}^{3} + v u_{1}^{3} = \omega^2 u_{0}^{1}
\end{equation}
\begin{equation}
-\left ( 2w+2v \right ) u_{0}^{2} + w u_{0}^{3} + w u_{0}^{1} + v u_{2}^{3} + v u_{3}^{1} = \omega^2 u_{0}^{2}
\end{equation}
\begin{equation}
-\left ( 2w+v \right ) u_{0}^{3} + w u_{0}^{1} + w u_{0}^{2} + v u_{4}^{1} +  = \omega^2 u_{0}^{3}
\end{equation}
\label{eq_6}
\end{subequations}
while the soft boundary case is the same as the form in Eq.\hyperref[eq_3]{(3)}.

Comparing Eq.\hyperref[eq_6]{(6)} with Eq.\hyperref[eq_3]{(3)}, it is obvious to see that the impacts of boundary condition exactly reflect in the difference of the diagonal terms in the Hamiltonian. Overall, the soft boundary holds that the diagonal terms of all lattices are equivalent as $-2w-2v$ in both bulk and boundary lattices. From this perspective, by applying soft boundaries, the closed systems can be considered as in the thermodynamic limit.

As discussed above, distinct from the structures with periodic boundaries, the open boundaries of the structures result in reduced symmetries in the boundary lattices. Next, we present the impacts of boundary conditions on the existence of topological states of the finite structure.

\section{\label{sec_4}boundary condition on topological states}
\begin{figure}[htbp]
\centering
\includegraphics[width=15cm]{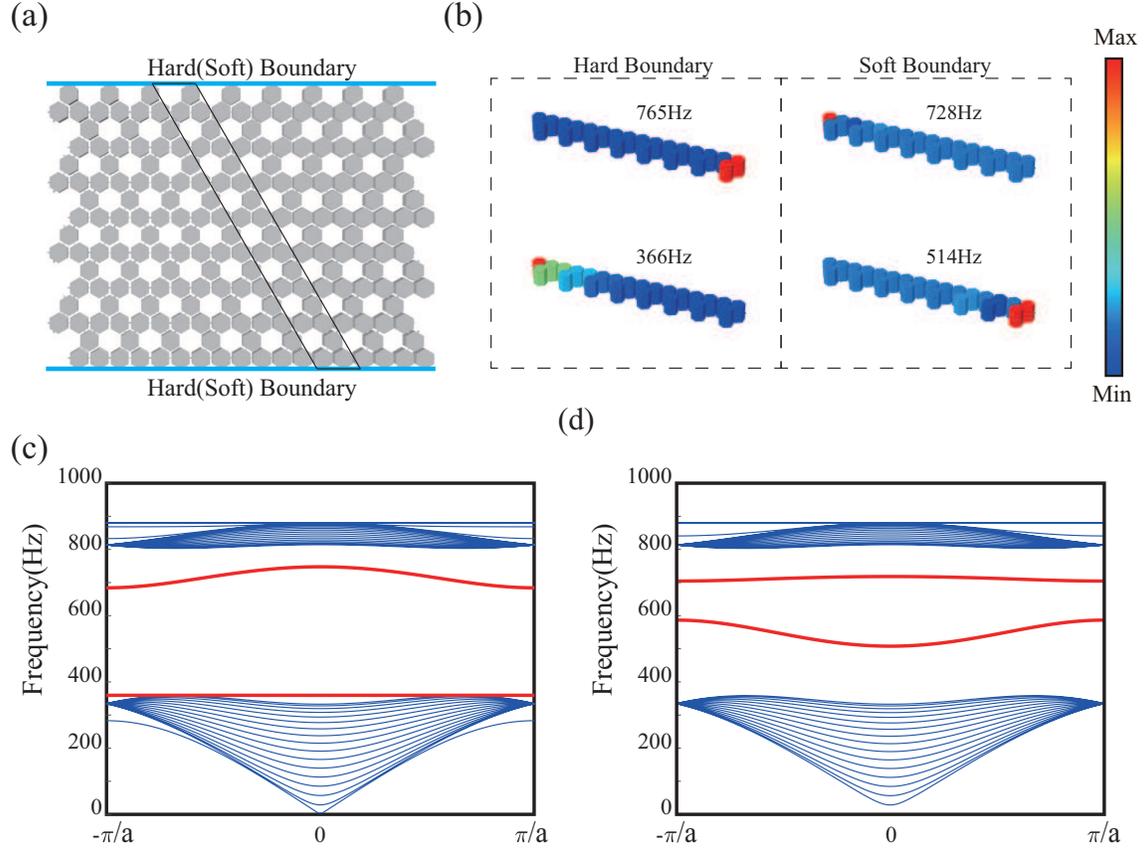}
\caption{(a) Schematic of the ribbon-shaped superlattice. The ribbon is extended along x-direction and enclosed by hard or soft boundary (marked with blue lines) in y-direction. (b) Sound pressure field distributions of the edge states at $k=0$. (c) The energy band structure of the superlattice with hard boundaries. (d) The energy band structure of the superlattice with soft boundaries.}
\label{fig_3}
\end{figure}

In this section, we first discuss the semi-infinite case. As shown in Fig. \hyperref[fig_3]{3(a)}, the ribbon-shaped superlattice that consists of 7 nontrivial kagome lattice with $v/w = 5$ being finite in y-direction, and is periodic in x-direction. It is worth noting that due to $ C_3 $ symmetry, the geometries of the upper edge and the lower edge are distinct \cite{ribbon}. Crucially, instead of employing interaction interfaces by lacking the nontrivial domain with so-called trivial “environment”, the upper edge and the lower edge in this work are connected only with hard or soft boundaries so that the y-direction is limited. Figures \hyperref[fig_3]{3(c)} and \hyperref[fig_3]{3(d)} depict the energy band structures of the superlattices with hard boundaries and hard boundaries, respectively, and their corresponding in-gap modes are presented in Fig. \hyperref[fig_2]{3(b)}. It is seen that the eigenfrequencies and the modes are various in different cases. For the hard boundary case, the modes of lower frequency locate in the upper boundary but the soft boundary opposites. Meanwhile, it is important to note that the upper in-gap mode in the hard boundary case is actually the ordinary resonance state induced by ordinary local resonance (though still labeled with red), while the lower in-gap modes are topological. This is due to the fact that the topological states are determined by the bulk properties, and the boundary properties only affect the corresponding energy of the states located at the edges. Detailedly, these isolated ordinary edge-localized states are separated from the upper bulk states, due to the hard boundary condition \cite{add1,add2}. As a result, by applying different boundary conditions, the existence of topological states is quite distinct.

\begin{figure}[htbp]
\centering
\includegraphics[width=15cm]{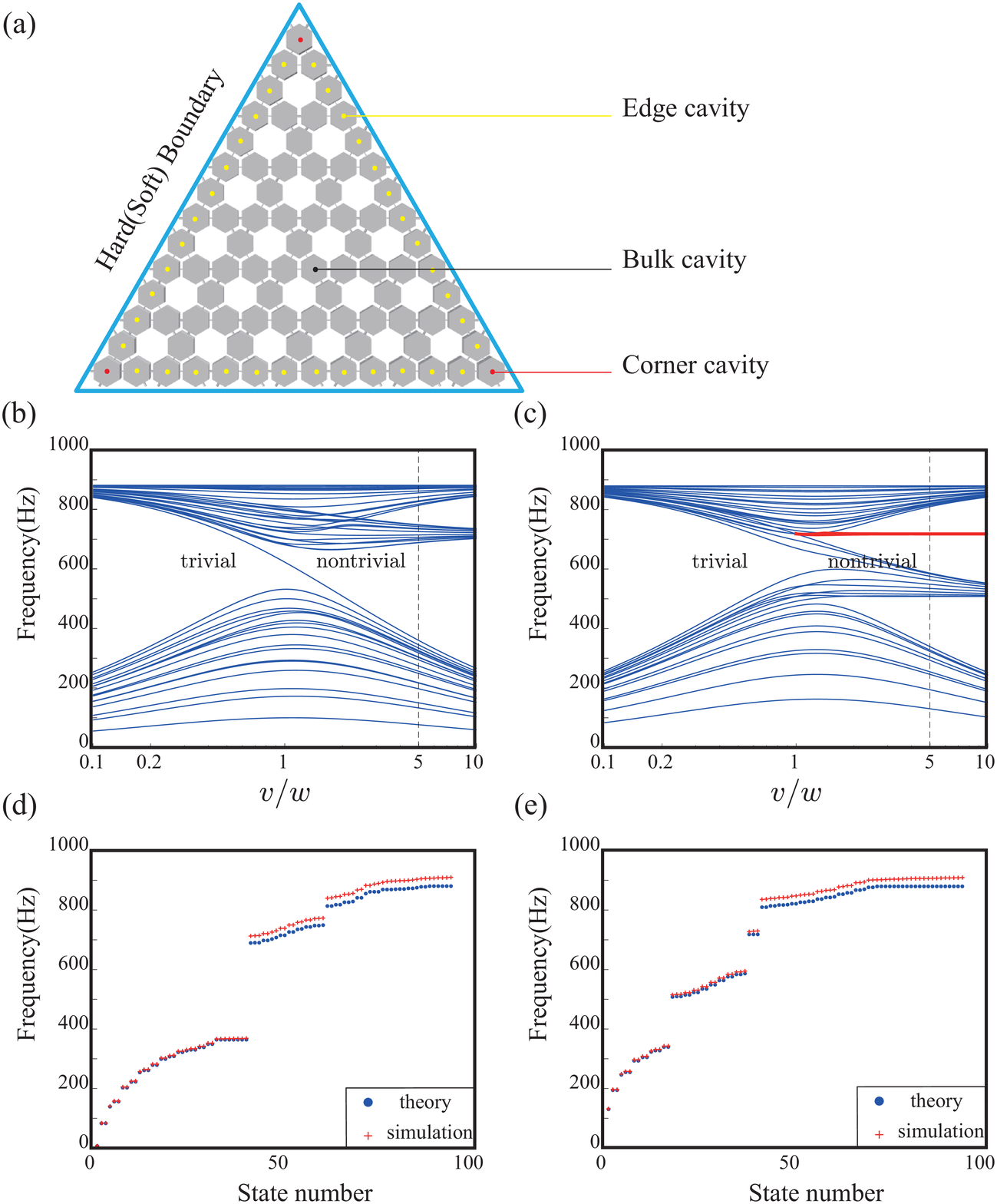}
\caption{(a) Schematic of the triangular finite structure consisting of kagome lattices. (b) Energy spectrum with hard boundary. (c) Energy spectrum with soft boundary. The corner state is emphasized with red line. (d) The eigenfrequencies when ${v}/{w}=5$ with hard boundary. (e) The calculated eigenfrequencies when ${v}/{w}=5$ with soft boundary. The theoretical frequency of corner state is 718Hz while the simulation one is 722Hz. The blue dots are theoretical results and the red ‘+’ marks are simulation results.}
\label{fig_4}
\end{figure}

Further, taking the outmost boundary condition into account, we demonstrate the existence of second-order topological states in two-dimensional materials with full open boundaries. Figure \hyperref[fig_4]{4(a)} exhibits the nontrivial triangle-shaped structure consisting of 28 kagome lattices, with the hard (soft) boundaries encloses the whole structure. Therefore, the lattices in the structure can be separately defined as bulk lattices, edge lattices and corner lattices. For the bulk lattices, there are six nearest-neighbor lattices, while there are four nearest-neighbors for the edge lattices and only two nearest-neighbors for the corner lattices, respectively. The relationship between the diagonal terms of the Hamiltonian and the boundary condition can be obtained through acoustic-electric analogy as presented Table. \hyperref[tab_1]{I}
\begin{table}[h]  
\caption{the diagonal terms of the Hamiltonian for different cavities} 
\begin{tabular}{p{2.5cm}|p{3cm}|p{3cm}}  
\hline  
\hline  
 & Hard & Soft \\ 
\hline   
Bulk cavities& $-2w-2v$ & $-2w-2v$\\  
\hline  
Edge cavities& $-2w-v$ & $-2w-2v$\\  
\hline  
Corner cavities& $-2w$ & $-2w-2v$\\   
\hline  
\hline  
\end{tabular}
\label{tab_1}
\end{table}

As a result, we can see that all the diagonal terms in Hamiltonian with the soft boundary are equivalent while the hard boundary ones are different, which meets the discussion in Sec. \hyperref[sec_3]{III}. Crucially, Figs. \hyperref[fig_4]{4(b)} and \hyperref[fig_4]{4(c)} show the processes of the change of band structure along with the ratio of $v/w$ in different boundary conditions. What stands out in the result is that only in the soft boundary condition the corner states can exist, which results from the preservation of $C_3$ symmetry and generalized chiral symmetry in contrast with hard boundary condition. Moreover, due to the equivalent on-site terms, the obtained Hamiltonian $\mathcal H_{0}$ in soft boundary condition is stringently corresponds to the theoretical model of TCI presented in condensed matter physics. Therefore, the induced corner states (labeled with red line in Fig. \hyperref[fig_4]{4(c)} exactly correspond to the zero-energy states. We calculate the frequency of corner state through follow equation:
\begin{equation}
f_{corner} = \frac{\sqrt{-2w-2v}}{2\pi} = \frac{1}{2\pi} \sqrt{\frac{2 \pi c^2}{V} \left ( \frac{r_v^2}{l_v + 1.7r_v} + \frac{r_w^2}{l_w + 1.7r_w} \right )}
\label{eq_8}
\end{equation}

\begin{figure}[htbp]
\centering
\includegraphics[width=12cm]{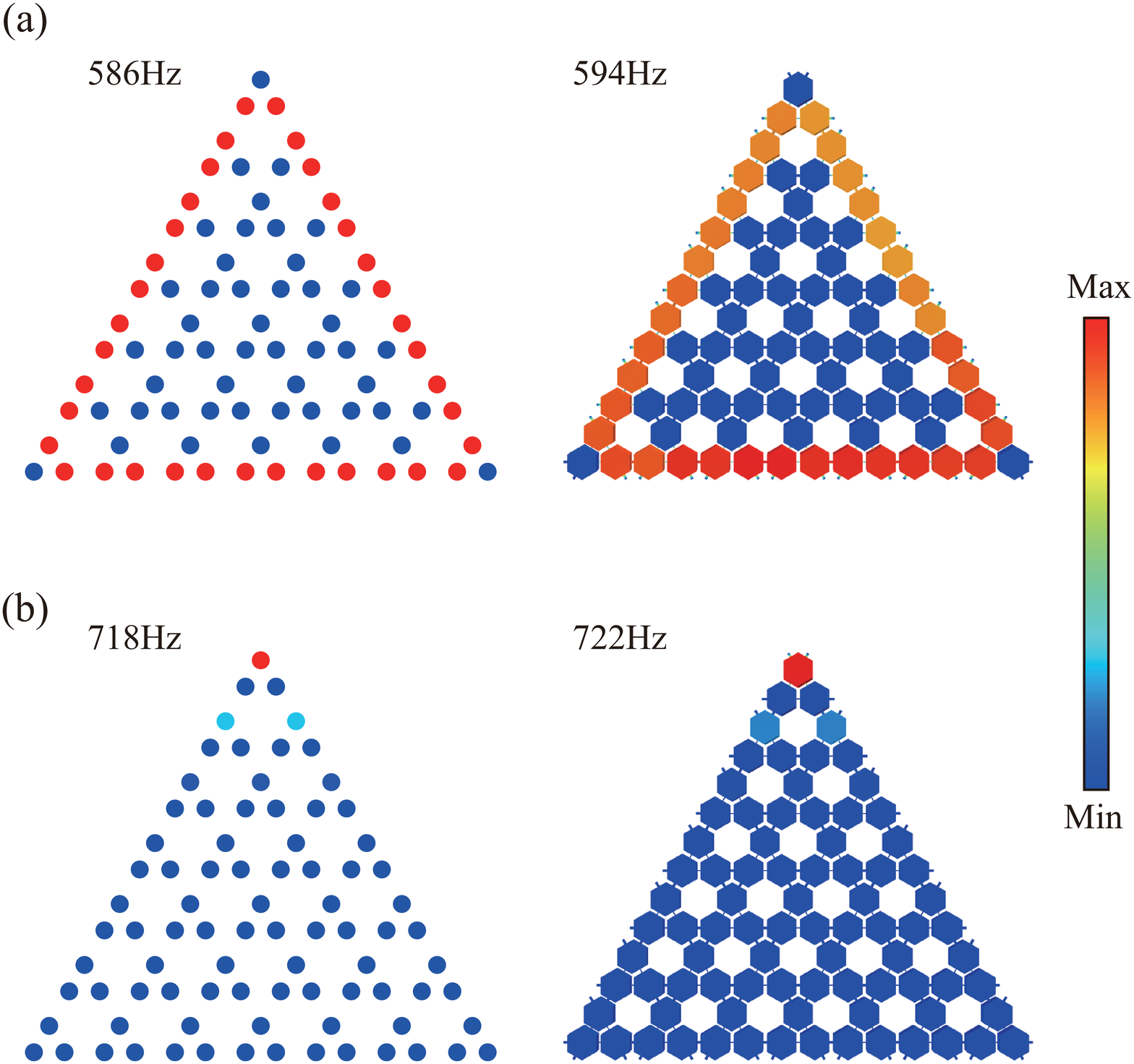}
\caption{Theoretical results and simulation results of edge state and corner state with soft boundary. (a) Sound pressure field distributions of the topological edge states. (b) Sound pressure field distributions of the topological corner states. The left panels are the theoretical results, and the right panels are the simulation results.}
\label{fig_5}
\end{figure}

Here, we set $v/w = 5$ to show our results. For comparison, the theoretical results and the simulation results are shown in Figs. \hyperref[fig_4]{4(d)} and \hyperref[fig_4]{4(e)}, respectively. The frequency of corner state with soft boundary condition is 718Hz in our calculation using Eq.\hyperref[eq_8]{(8)} and the simulation result is 727Hz. Meanwhile, the theoretical and numerical results of of edge state and the 'zero-energy' corner state are shown in Figs. \hyperref[fig_5]{5(a)} and \hyperref[fig_5]{5(b)}, respectively. 

\section{\label{sec_6}conlusion}
In conclusion, we have proposed a general way to obtain the rigorous Hamiltonian based on a acoustic TCI with a kagome lattice, which reveals the higher-order topology in acoustic systems. Through such the approach, the relationship between the diagonal terms of Hamiltonian and the boundary conditions for a finite structure is well expounded, which can crucially impact the topological states of closed systems. We demonstrate that only the soft boundary coincides with the Hamiltonian whose diagonal terms are zero in electronic system precisely, in which zero-energy corner states exist. Crucially, this approach in turn can construct the desired Hamiltonians that are with specific topological phenomena, and then directly design the precise physical structures. Meanwhile, we note that this method can easily be extended to higher dimensions. Our work is useful for the understanding of topology in acoustic systems, and is expected to provide a new general platform for the design of the topological acoustic materials quantitatively.

\begin{acknowledgments}
This work was supported by the National Key R\&D Program of China (Grant No. 2017YFA0303700), National Natural Science Foundation of China (Grant Nos. 11634006, 11934009, and 12074184), the Natural Science Foundation of Jiangsu Province (Grant No. BK20191245), State Key Laboratory of Acoustics, Chinese Academy of Sciences.
\end{acknowledgments}


\begin{thebibliography}{50}%
\makeatletter
\providecommand \@ifxundefined [1]{%
 \@ifx{#1\undefined}
}%
\providecommand \@ifnum [1]{%
 \ifnum #1\expandafter \@firstoftwo
 \else \expandafter \@secondoftwo
 \fi
}%
\providecommand \@ifx [1]{%
 \ifx #1\expandafter \@firstoftwo
 \else \expandafter \@secondoftwo
 \fi
}%
\providecommand \natexlab [1]{#1}%
\providecommand \enquote  [1]{``#1''}%
\providecommand \bibnamefont  [1]{#1}%
\providecommand \bibfnamefont [1]{#1}%
\providecommand \citenamefont [1]{#1}%
\providecommand \href@noop [0]{\@secondoftwo}%
\providecommand \href [0]{\begingroup \@sanitize@url \@href}%
\providecommand \@href[1]{\@@startlink{#1}\@@href}%
\providecommand \@@href[1]{\endgroup#1\@@endlink}%
\providecommand \@sanitize@url [0]{\catcode `\\12\catcode `\$12\catcode
  `\&12\catcode `\#12\catcode `\^12\catcode `\_12\catcode `\%12\relax}%
\providecommand \@@startlink[1]{}%
\providecommand \@@endlink[0]{}%
\providecommand \url  [0]{\begingroup\@sanitize@url \@url }%
\providecommand \@url [1]{\endgroup\@href {#1}{\urlprefix }}%
\providecommand \urlprefix  [0]{URL }%
\providecommand \Eprint [0]{\href }%
\providecommand \doibase [0]{https://doi.org/}%
\providecommand \selectlanguage [0]{\@gobble}%
\providecommand \bibinfo  [0]{\@secondoftwo}%
\providecommand \bibfield  [0]{\@secondoftwo}%
\providecommand \translation [1]{[#1]}%
\providecommand \BibitemOpen [0]{}%
\providecommand \bibitemStop [0]{}%
\providecommand \bibitemNoStop [0]{.\EOS\space}%
\providecommand \EOS [0]{\spacefactor3000\relax}%
\providecommand \BibitemShut  [1]{\csname bibitem#1\endcsname}%
\let\auto@bib@innerbib\@empty
\bibitem [{\citenamefont {Hasan}\ and\ \citenamefont {Kane}(2010)}]{cm_1}%
  \BibitemOpen
  \bibfield  {author} {\bibinfo {author} {\bibfnamefont {M.~Z.}\ \bibnamefont
  {Hasan}}\ and\ \bibinfo {author} {\bibfnamefont {C.~L.}\ \bibnamefont
  {Kane}},\ }\bibfield  {title} {\bibinfo {title} {Colloquium: Topological
  insulators},\ }\href {https://doi.org/10.1103/revmodphys.82.3045} {\bibfield
  {journal} {\bibinfo  {journal} {Reviews of Modern Physics}\ }\textbf
  {\bibinfo {volume} {82}},\ \bibinfo {pages} {3045} (\bibinfo {year}
  {2010})}\BibitemShut {NoStop}%
\bibitem [{\citenamefont {Qi}\ and\ \citenamefont {Zhang}(2011)}]{cm_2}%
  \BibitemOpen
  \bibfield  {author} {\bibinfo {author} {\bibfnamefont {X.-L.}\ \bibnamefont
  {Qi}}\ and\ \bibinfo {author} {\bibfnamefont {S.-C.}\ \bibnamefont {Zhang}},\
  }\bibfield  {title} {\bibinfo {title} {Topological insulators and
  superconductors},\ }\href {https://doi.org/10.1103/revmodphys.83.1057}
  {\bibfield  {journal} {\bibinfo  {journal} {Reviews of Modern Physics}\
  }\textbf {\bibinfo {volume} {83}},\ \bibinfo {pages} {1057} (\bibinfo {year}
  {2011})}\BibitemShut {NoStop}%
\bibitem [{\citenamefont {Haldane}(1988)}]{cm_3}%
  \BibitemOpen
  \bibfield  {author} {\bibinfo {author} {\bibfnamefont {F.~D.~M.}\
  \bibnamefont {Haldane}},\ }\bibfield  {title} {\bibinfo {title} {Model for a
  quantum hall effect without landau levels: Condensed-matter realization of
  the "parity anomaly"},\ }\href {https://doi.org/10.1103/physrevlett.61.2015}
  {\bibfield  {journal} {\bibinfo  {journal} {Physical Review Letters}\
  }\textbf {\bibinfo {volume} {61}},\ \bibinfo {pages} {2015} (\bibinfo {year}
  {1988})}\BibitemShut {NoStop}%
\bibitem [{\citenamefont {Zhang}\ \emph
  {et~al.}(2020{\natexlab{a}})\citenamefont {Zhang}, \citenamefont {Chen},
  \citenamefont {Zhang},\ and\ \citenamefont {Hu}}]{cw_1}%
  \BibitemOpen
  \bibfield  {author} {\bibinfo {author} {\bibfnamefont {Q.}~\bibnamefont
  {Zhang}}, \bibinfo {author} {\bibfnamefont {Y.}~\bibnamefont {Chen}},
  \bibinfo {author} {\bibfnamefont {K.}~\bibnamefont {Zhang}},\ and\ \bibinfo
  {author} {\bibfnamefont {G.}~\bibnamefont {Hu}},\ }\bibfield  {title}
  {\bibinfo {title} {Dirac degeneracy and elastic topological valley modes
  induced by local resonant states},\ }\href
  {https://doi.org/10.1103/physrevb.101.014101} {\bibfield  {journal} {\bibinfo
   {journal} {Physical Review B}\ }\textbf {\bibinfo {volume} {101}},\ \bibinfo
  {pages} {014101} (\bibinfo {year} {2020}{\natexlab{a}})}\BibitemShut
  {NoStop}%
\bibitem [{\citenamefont {Zhang}\ \emph {et~al.}(2019)\citenamefont {Zhang},
  \citenamefont {Wang}, \citenamefont {Lin}, \citenamefont {Tian},
  \citenamefont {Xie}, \citenamefont {Lu}, \citenamefont {Chen},\ and\
  \citenamefont {Jiang}}]{cw_2}%
  \BibitemOpen
  \bibfield  {author} {\bibinfo {author} {\bibfnamefont {X.}~\bibnamefont
  {Zhang}}, \bibinfo {author} {\bibfnamefont {H.-X.}\ \bibnamefont {Wang}},
  \bibinfo {author} {\bibfnamefont {Z.-K.}\ \bibnamefont {Lin}}, \bibinfo
  {author} {\bibfnamefont {Y.}~\bibnamefont {Tian}}, \bibinfo {author}
  {\bibfnamefont {B.}~\bibnamefont {Xie}}, \bibinfo {author} {\bibfnamefont
  {M.-H.}\ \bibnamefont {Lu}}, \bibinfo {author} {\bibfnamefont {Y.-F.}\
  \bibnamefont {Chen}},\ and\ \bibinfo {author} {\bibfnamefont {J.-H.}\
  \bibnamefont {Jiang}},\ }\bibfield  {title} {\bibinfo {title} {Second-order
  topology and multidimensional topological transitions in sonic crystals},\
  }\href {https://doi.org/10.1038/s41567-019-0472-1} {\bibfield  {journal}
  {\bibinfo  {journal} {Nature Physics}\ }\textbf {\bibinfo {volume} {15}},\
  \bibinfo {pages} {582} (\bibinfo {year} {2019})}\BibitemShut {NoStop}%
\bibitem [{\citenamefont {Wang}\ \emph {et~al.}(2019)\citenamefont {Wang},
  \citenamefont {Guo},\ and\ \citenamefont {Jiang}}]{cw_3}%
  \BibitemOpen
  \bibfield  {author} {\bibinfo {author} {\bibfnamefont {H.-X.}\ \bibnamefont
  {Wang}}, \bibinfo {author} {\bibfnamefont {G.-Y.}\ \bibnamefont {Guo}},\ and\
  \bibinfo {author} {\bibfnamefont {J.-H.}\ \bibnamefont {Jiang}},\ }\bibfield
  {title} {\bibinfo {title} {Band topology in classical waves: Wilson-loop
  approach to topological numbers and fragile topology},\ }\href
  {https://doi.org/10.1088/1367-2630/ab3f71} {\bibfield  {journal} {\bibinfo
  {journal} {New Journal of Physics}\ }\textbf {\bibinfo {volume} {21}},\
  \bibinfo {pages} {093029} (\bibinfo {year} {2019})}\BibitemShut {NoStop}%
\bibitem [{\citenamefont {Chen}\ \emph
  {et~al.}(2019{\natexlab{a}})\citenamefont {Chen}, \citenamefont {Liu},\ and\
  \citenamefont {Hu}}]{cw_4}%
  \BibitemOpen
  \bibfield  {author} {\bibinfo {author} {\bibfnamefont {Y.}~\bibnamefont
  {Chen}}, \bibinfo {author} {\bibfnamefont {X.}~\bibnamefont {Liu}},\ and\
  \bibinfo {author} {\bibfnamefont {G.}~\bibnamefont {Hu}},\ }\bibfield
  {title} {\bibinfo {title} {Topological phase transition in mechanical
  honeycomb lattice},\ }\href {https://doi.org/10.1016/j.jmps.2018.08.021}
  {\bibfield  {journal} {\bibinfo  {journal} {Journal of the Mechanics and
  Physics of Solids}\ }\textbf {\bibinfo {volume} {122}},\ \bibinfo {pages}
  {54} (\bibinfo {year} {2019}{\natexlab{a}})}\BibitemShut {NoStop}%
\bibitem [{\citenamefont {Xue}\ \emph {et~al.}(2018)\citenamefont {Xue},
  \citenamefont {Yang}, \citenamefont {Gao}, \citenamefont {Chong},\ and\
  \citenamefont {Zhang}}]{cw_5}%
  \BibitemOpen
  \bibfield  {author} {\bibinfo {author} {\bibfnamefont {H.}~\bibnamefont
  {Xue}}, \bibinfo {author} {\bibfnamefont {Y.}~\bibnamefont {Yang}}, \bibinfo
  {author} {\bibfnamefont {F.}~\bibnamefont {Gao}}, \bibinfo {author}
  {\bibfnamefont {Y.}~\bibnamefont {Chong}},\ and\ \bibinfo {author}
  {\bibfnamefont {B.}~\bibnamefont {Zhang}},\ }\bibfield  {title} {\bibinfo
  {title} {Acoustic higher-order topological insulator on a kagome lattice},\
  }\href {https://doi.org/10.1038/s41563-018-0251-x} {\bibfield  {journal}
  {\bibinfo  {journal} {Nature Materials}\ }\textbf {\bibinfo {volume} {18}},\
  \bibinfo {pages} {108} (\bibinfo {year} {2018})}\BibitemShut {NoStop}%
\bibitem [{\citenamefont {Serra-Garcia}\ \emph {et~al.}(2018)\citenamefont
  {Serra-Garcia}, \citenamefont {Peri}, \citenamefont {Süsstrunk},
  \citenamefont {Bilal}, \citenamefont {Larsen}, \citenamefont {Villanueva},\
  and\ \citenamefont {Huber}}]{cw_6}%
  \BibitemOpen
  \bibfield  {author} {\bibinfo {author} {\bibfnamefont {M.}~\bibnamefont
  {Serra-Garcia}}, \bibinfo {author} {\bibfnamefont {V.}~\bibnamefont {Peri}},
  \bibinfo {author} {\bibfnamefont {R.}~\bibnamefont {Süsstrunk}}, \bibinfo
  {author} {\bibfnamefont {O.~R.}\ \bibnamefont {Bilal}}, \bibinfo {author}
  {\bibfnamefont {T.}~\bibnamefont {Larsen}}, \bibinfo {author} {\bibfnamefont
  {L.~G.}\ \bibnamefont {Villanueva}},\ and\ \bibinfo {author} {\bibfnamefont
  {S.~D.}\ \bibnamefont {Huber}},\ }\bibfield  {title} {\bibinfo {title}
  {Observation of a phononic quadrupole topological insulator},\ }\href
  {https://doi.org/10.1038/nature25156} {\bibfield  {journal} {\bibinfo
  {journal} {Nature}\ }\textbf {\bibinfo {volume} {555}},\ \bibinfo {pages}
  {342} (\bibinfo {year} {2018})}\BibitemShut {NoStop}%
\bibitem [{\citenamefont {Noh}\ \emph {et~al.}(2018)\citenamefont {Noh},
  \citenamefont {Benalcazar}, \citenamefont {Huang}, \citenamefont {Collins},
  \citenamefont {Chen}, \citenamefont {Hughes},\ and\ \citenamefont
  {Rechtsman}}]{cw_7}%
  \BibitemOpen
  \bibfield  {author} {\bibinfo {author} {\bibfnamefont {J.}~\bibnamefont
  {Noh}}, \bibinfo {author} {\bibfnamefont {W.~A.}\ \bibnamefont {Benalcazar}},
  \bibinfo {author} {\bibfnamefont {S.}~\bibnamefont {Huang}}, \bibinfo
  {author} {\bibfnamefont {M.~J.}\ \bibnamefont {Collins}}, \bibinfo {author}
  {\bibfnamefont {K.~P.}\ \bibnamefont {Chen}}, \bibinfo {author}
  {\bibfnamefont {T.~L.}\ \bibnamefont {Hughes}},\ and\ \bibinfo {author}
  {\bibfnamefont {M.~C.}\ \bibnamefont {Rechtsman}},\ }\bibfield  {title}
  {\bibinfo {title} {Topological protection of photonic mid-gap defect modes},\
  }\href {https://doi.org/10.1038/s41566-018-0179-3} {\bibfield  {journal}
  {\bibinfo  {journal} {Nature Photonics}\ }\textbf {\bibinfo {volume} {12}},\
  \bibinfo {pages} {408} (\bibinfo {year} {2018})}\BibitemShut {NoStop}%
\bibitem [{\citenamefont {Ni}\ \emph {et~al.}(2018)\citenamefont {Ni},
  \citenamefont {Weiner}, \citenamefont {Alù},\ and\ \citenamefont
  {Khanikaev}}]{cw_8}%
  \BibitemOpen
  \bibfield  {author} {\bibinfo {author} {\bibfnamefont {X.}~\bibnamefont
  {Ni}}, \bibinfo {author} {\bibfnamefont {M.}~\bibnamefont {Weiner}}, \bibinfo
  {author} {\bibfnamefont {A.}~\bibnamefont {Alù}},\ and\ \bibinfo {author}
  {\bibfnamefont {A.~B.}\ \bibnamefont {Khanikaev}},\ }\bibfield  {title}
  {\bibinfo {title} {Observation of higher-order topological acoustic states
  protected by generalized chiral symmetry},\ }\href
  {https://doi.org/10.1038/s41563-018-0252-9} {\bibfield  {journal} {\bibinfo
  {journal} {Nature Materials}\ }\textbf {\bibinfo {volume} {18}},\ \bibinfo
  {pages} {113} (\bibinfo {year} {2018})}\BibitemShut {NoStop}%
\bibitem [{\citenamefont {Peng}\ \emph {et~al.}(2016)\citenamefont {Peng},
  \citenamefont {Qin}, \citenamefont {Zhao}, \citenamefont {Shen},
  \citenamefont {Xu}, \citenamefont {Bao}, \citenamefont {Jia},\ and\
  \citenamefont {Zhu}}]{cw_9}%
  \BibitemOpen
  \bibfield  {author} {\bibinfo {author} {\bibfnamefont {Y.-G.}\ \bibnamefont
  {Peng}}, \bibinfo {author} {\bibfnamefont {C.-Z.}\ \bibnamefont {Qin}},
  \bibinfo {author} {\bibfnamefont {D.-G.}\ \bibnamefont {Zhao}}, \bibinfo
  {author} {\bibfnamefont {Y.-X.}\ \bibnamefont {Shen}}, \bibinfo {author}
  {\bibfnamefont {X.-Y.}\ \bibnamefont {Xu}}, \bibinfo {author} {\bibfnamefont
  {M.}~\bibnamefont {Bao}}, \bibinfo {author} {\bibfnamefont {H.}~\bibnamefont
  {Jia}},\ and\ \bibinfo {author} {\bibfnamefont {X.-F.}\ \bibnamefont {Zhu}},\
  }\bibfield  {title} {\bibinfo {title} {Experimental demonstration of
  anomalous floquet topological insulator for sound},\ }\href
  {https://doi.org/10.1038/ncomms13368} {\bibfield  {journal} {\bibinfo
  {journal} {Nature Communications}\ }\textbf {\bibinfo {volume} {7}},\
  \bibinfo {pages} {13368} (\bibinfo {year} {2016})}\BibitemShut {NoStop}%
\bibitem [{\citenamefont {He}\ \emph {et~al.}(2016)\citenamefont {He},
  \citenamefont {Ni}, \citenamefont {Ge}, \citenamefont {Sun}, \citenamefont
  {Chen}, \citenamefont {Lu}, \citenamefont {Liu},\ and\ \citenamefont
  {Chen}}]{cw_10}%
  \BibitemOpen
  \bibfield  {author} {\bibinfo {author} {\bibfnamefont {C.}~\bibnamefont
  {He}}, \bibinfo {author} {\bibfnamefont {X.}~\bibnamefont {Ni}}, \bibinfo
  {author} {\bibfnamefont {H.}~\bibnamefont {Ge}}, \bibinfo {author}
  {\bibfnamefont {X.-C.}\ \bibnamefont {Sun}}, \bibinfo {author} {\bibfnamefont
  {Y.-B.}\ \bibnamefont {Chen}}, \bibinfo {author} {\bibfnamefont {M.-H.}\
  \bibnamefont {Lu}}, \bibinfo {author} {\bibfnamefont {X.-P.}\ \bibnamefont
  {Liu}},\ and\ \bibinfo {author} {\bibfnamefont {Y.-F.}\ \bibnamefont
  {Chen}},\ }\bibfield  {title} {\bibinfo {title} {Acoustic topological
  insulator and robust one-way sound transport},\ }\href
  {https://doi.org/10.1038/nphys3867} {\bibfield  {journal} {\bibinfo
  {journal} {Nature Physics}\ }\textbf {\bibinfo {volume} {12}},\ \bibinfo
  {pages} {1124} (\bibinfo {year} {2016})}\BibitemShut {NoStop}%
\bibitem [{\citenamefont {Yang}\ \emph {et~al.}(2015)\citenamefont {Yang},
  \citenamefont {Gao}, \citenamefont {Shi}, \citenamefont {Lin}, \citenamefont
  {Gao}, \citenamefont {Chong},\ and\ \citenamefont {Zhang}}]{cw_11}%
  \BibitemOpen
  \bibfield  {author} {\bibinfo {author} {\bibfnamefont {Z.}~\bibnamefont
  {Yang}}, \bibinfo {author} {\bibfnamefont {F.}~\bibnamefont {Gao}}, \bibinfo
  {author} {\bibfnamefont {X.}~\bibnamefont {Shi}}, \bibinfo {author}
  {\bibfnamefont {X.}~\bibnamefont {Lin}}, \bibinfo {author} {\bibfnamefont
  {Z.}~\bibnamefont {Gao}}, \bibinfo {author} {\bibfnamefont {Y.}~\bibnamefont
  {Chong}},\ and\ \bibinfo {author} {\bibfnamefont {B.}~\bibnamefont {Zhang}},\
  }\bibfield  {title} {\bibinfo {title} {Topological acoustics},\ }\href
  {https://doi.org/10.1103/physrevlett.114.114301} {\bibfield  {journal}
  {\bibinfo  {journal} {Physical Review Letters}\ }\textbf {\bibinfo {volume}
  {114}},\ \bibinfo {pages} {114301} (\bibinfo {year} {2015})}\BibitemShut
  {NoStop}%
\bibitem [{\citenamefont {Xiao}\ \emph {et~al.}(2015)\citenamefont {Xiao},
  \citenamefont {Ma}, \citenamefont {Yang}, \citenamefont {Sheng},
  \citenamefont {Zhang},\ and\ \citenamefont {Chan}}]{cw_12}%
  \BibitemOpen
  \bibfield  {author} {\bibinfo {author} {\bibfnamefont {M.}~\bibnamefont
  {Xiao}}, \bibinfo {author} {\bibfnamefont {G.}~\bibnamefont {Ma}}, \bibinfo
  {author} {\bibfnamefont {Z.}~\bibnamefont {Yang}}, \bibinfo {author}
  {\bibfnamefont {P.}~\bibnamefont {Sheng}}, \bibinfo {author} {\bibfnamefont
  {Z.~Q.}\ \bibnamefont {Zhang}},\ and\ \bibinfo {author} {\bibfnamefont
  {C.~T.}\ \bibnamefont {Chan}},\ }\bibfield  {title} {\bibinfo {title}
  {Geometric phase and band inversion in periodic acoustic systems},\ }\href
  {https://doi.org/10.1038/nphys3228} {\bibfield  {journal} {\bibinfo
  {journal} {Nature Physics}\ }\textbf {\bibinfo {volume} {11}},\ \bibinfo
  {pages} {240} (\bibinfo {year} {2015})}\BibitemShut {NoStop}%
\bibitem [{\citenamefont {Chen}\ \emph {et~al.}(2014)\citenamefont {Chen},
  \citenamefont {Jiang}, \citenamefont {Chen}, \citenamefont {Zhu},
  \citenamefont {Zhou}, \citenamefont {Dong},\ and\ \citenamefont
  {Chan}}]{cw_13}%
  \BibitemOpen
  \bibfield  {author} {\bibinfo {author} {\bibfnamefont {W.-J.}\ \bibnamefont
  {Chen}}, \bibinfo {author} {\bibfnamefont {S.-J.}\ \bibnamefont {Jiang}},
  \bibinfo {author} {\bibfnamefont {X.-D.}\ \bibnamefont {Chen}}, \bibinfo
  {author} {\bibfnamefont {B.}~\bibnamefont {Zhu}}, \bibinfo {author}
  {\bibfnamefont {L.}~\bibnamefont {Zhou}}, \bibinfo {author} {\bibfnamefont
  {J.-W.}\ \bibnamefont {Dong}},\ and\ \bibinfo {author} {\bibfnamefont
  {C.~T.}\ \bibnamefont {Chan}},\ }\bibfield  {title} {\bibinfo {title}
  {Experimental realization of photonic topological insulator in a uniaxial
  metacrystal waveguide},\ }\href {https://doi.org/10.1038/ncomms6782}
  {\bibfield  {journal} {\bibinfo  {journal} {Nature Communications}\ }\textbf
  {\bibinfo {volume} {5}},\ \bibinfo {pages} {5782} (\bibinfo {year}
  {2014})}\BibitemShut {NoStop}%
\bibitem [{\citenamefont {Wang}\ \emph {et~al.}(2008)\citenamefont {Wang},
  \citenamefont {Chong}, \citenamefont {Joannopoulos},\ and\ \citenamefont
  {Soljačić}}]{cw_14}%
  \BibitemOpen
  \bibfield  {author} {\bibinfo {author} {\bibfnamefont {Z.}~\bibnamefont
  {Wang}}, \bibinfo {author} {\bibfnamefont {Y.~D.}\ \bibnamefont {Chong}},
  \bibinfo {author} {\bibfnamefont {J.~D.}\ \bibnamefont {Joannopoulos}},\ and\
  \bibinfo {author} {\bibfnamefont {M.}~\bibnamefont {Soljačić}},\ }\bibfield
   {title} {\bibinfo {title} {Reflection-free one-way edge modes in a
  gyromagnetic photonic crystal},\ }\href
  {https://doi.org/10.1103/physrevlett.100.013905} {\bibfield  {journal}
  {\bibinfo  {journal} {Physical Review Letters}\ }\textbf {\bibinfo {volume}
  {100}},\ \bibinfo {pages} {013905} (\bibinfo {year} {2008})}\BibitemShut
  {NoStop}%
\bibitem [{\citenamefont {Deng}\ \emph {et~al.}(2020)\citenamefont {Deng},
  \citenamefont {Huang}, \citenamefont {Lu}, \citenamefont {Peri},
  \citenamefont {Li}, \citenamefont {Huber},\ and\ \citenamefont
  {Liu}}]{Deng1}%
  \BibitemOpen
  \bibfield  {author} {\bibinfo {author} {\bibfnamefont {W.}~\bibnamefont
  {Deng}}, \bibinfo {author} {\bibfnamefont {X.}~\bibnamefont {Huang}},
  \bibinfo {author} {\bibfnamefont {J.}~\bibnamefont {Lu}}, \bibinfo {author}
  {\bibfnamefont {V.}~\bibnamefont {Peri}}, \bibinfo {author} {\bibfnamefont
  {F.}~\bibnamefont {Li}}, \bibinfo {author} {\bibfnamefont {S.~D.}\
  \bibnamefont {Huber}},\ and\ \bibinfo {author} {\bibfnamefont
  {Z.}~\bibnamefont {Liu}},\ }\bibfield  {title} {\bibinfo {title} {Acoustic
  spin-chern insulator induced by synthetic spin–orbit coupling with spin
  conservation breaking},\ }\href {https://doi.org/10.1038/s41467-020-17039-1}
  {\bibfield  {journal} {\bibinfo  {journal} {Nat. Commun.}\ }\textbf {\bibinfo
  {volume} {11}},\ \bibinfo {pages} {3227} (\bibinfo {year}
  {2020})}\BibitemShut {NoStop}%
\bibitem [{\citenamefont {Haldane}\ and\ \citenamefont {Raghu}(2008)}]{cw_15}%
  \BibitemOpen
  \bibfield  {author} {\bibinfo {author} {\bibfnamefont {F.~D.~M.}\
  \bibnamefont {Haldane}}\ and\ \bibinfo {author} {\bibfnamefont
  {S.}~\bibnamefont {Raghu}},\ }\bibfield  {title} {\bibinfo {title} {Possible
  realization of directional optical waveguides in photonic crystals with
  broken time-reversal symmetry},\ }\href
  {https://doi.org/10.1103/physrevlett.100.013904} {\bibfield  {journal}
  {\bibinfo  {journal} {Physical Review Letters}\ }\textbf {\bibinfo {volume}
  {100}},\ \bibinfo {pages} {013904} (\bibinfo {year} {2008})}\BibitemShut
  {NoStop}%
\bibitem [{\citenamefont {Imhof}\ \emph {et~al.}(2018)\citenamefont {Imhof},
  \citenamefont {Berger}, \citenamefont {Bayer}, \citenamefont {Brehm},
  \citenamefont {Molenkamp}, \citenamefont {Kiessling}, \citenamefont
  {Schindler}, \citenamefont {Lee}, \citenamefont {Greiter}, \citenamefont
  {Neupert},\ and\ \citenamefont {Thomale}}]{cw_16}%
  \BibitemOpen
  \bibfield  {author} {\bibinfo {author} {\bibfnamefont {S.}~\bibnamefont
  {Imhof}}, \bibinfo {author} {\bibfnamefont {C.}~\bibnamefont {Berger}},
  \bibinfo {author} {\bibfnamefont {F.}~\bibnamefont {Bayer}}, \bibinfo
  {author} {\bibfnamefont {J.}~\bibnamefont {Brehm}}, \bibinfo {author}
  {\bibfnamefont {L.~W.}\ \bibnamefont {Molenkamp}}, \bibinfo {author}
  {\bibfnamefont {T.}~\bibnamefont {Kiessling}}, \bibinfo {author}
  {\bibfnamefont {F.}~\bibnamefont {Schindler}}, \bibinfo {author}
  {\bibfnamefont {C.~H.}\ \bibnamefont {Lee}}, \bibinfo {author} {\bibfnamefont
  {M.}~\bibnamefont {Greiter}}, \bibinfo {author} {\bibfnamefont
  {T.}~\bibnamefont {Neupert}},\ and\ \bibinfo {author} {\bibfnamefont
  {R.}~\bibnamefont {Thomale}},\ }\bibfield  {title} {\bibinfo {title}
  {Topolectrical-circuit realization of topological corner modes},\ }\href
  {https://doi.org/10.1038/s41567-018-0246-1} {\bibfield  {journal} {\bibinfo
  {journal} {Nature Physics}\ }\textbf {\bibinfo {volume} {14}},\ \bibinfo
  {pages} {925} (\bibinfo {year} {2018})}\BibitemShut {NoStop}%
\bibitem [{\citenamefont {Zhang}\ \emph
  {et~al.}(2020{\natexlab{b}})\citenamefont {Zhang}, \citenamefont {Hu},
  \citenamefont {Liu}, \citenamefont {Cheng}, \citenamefont {Liu},\ and\
  \citenamefont {Christensen}}]{cw_17}%
  \BibitemOpen
  \bibfield  {author} {\bibinfo {author} {\bibfnamefont {Z.}~\bibnamefont
  {Zhang}}, \bibinfo {author} {\bibfnamefont {B.}~\bibnamefont {Hu}}, \bibinfo
  {author} {\bibfnamefont {F.}~\bibnamefont {Liu}}, \bibinfo {author}
  {\bibfnamefont {Y.}~\bibnamefont {Cheng}}, \bibinfo {author} {\bibfnamefont
  {X.}~\bibnamefont {Liu}},\ and\ \bibinfo {author} {\bibfnamefont
  {J.}~\bibnamefont {Christensen}},\ }\bibfield  {title} {\bibinfo {title}
  {Pseudospin induced topological corner state at intersecting sonic
  lattices},\ }\href {https://doi.org/10.1103/PhysRevB.101.220102} {\bibfield
  {journal} {\bibinfo  {journal} {Physical Review B}\ }\textbf {\bibinfo
  {volume} {101}},\ \bibinfo {pages} {220102} (\bibinfo {year}
  {2020}{\natexlab{b}})}\BibitemShut {NoStop}%
\bibitem [{\citenamefont {Fan}\ \emph {et~al.}(2019)\citenamefont {Fan},
  \citenamefont {Xia}, \citenamefont {Tong}, \citenamefont {Zheng},\ and\
  \citenamefont {Yu}}]{cw_18}%
  \BibitemOpen
  \bibfield  {author} {\bibinfo {author} {\bibfnamefont {H.}~\bibnamefont
  {Fan}}, \bibinfo {author} {\bibfnamefont {B.}~\bibnamefont {Xia}}, \bibinfo
  {author} {\bibfnamefont {L.}~\bibnamefont {Tong}}, \bibinfo {author}
  {\bibfnamefont {S.}~\bibnamefont {Zheng}},\ and\ \bibinfo {author}
  {\bibfnamefont {D.}~\bibnamefont {Yu}},\ }\bibfield  {title} {\bibinfo
  {title} {Elastic higher-order topological insulator with topologically
  protected corner states},\ }\href
  {https://doi.org/10.1103/PhysRevLett.122.204301} {\bibfield  {journal}
  {\bibinfo  {journal} {Physical Review Letters}\ }\textbf {\bibinfo {volume}
  {122}},\ \bibinfo {pages} {204301} (\bibinfo {year} {2019})}\BibitemShut
  {NoStop}%
\bibitem [{\citenamefont {Hughes}\ \emph {et~al.}(2011)\citenamefont {Hughes},
  \citenamefont {Prodan},\ and\ \citenamefont {Bernevig}}]{tci_2}%
  \BibitemOpen
  \bibfield  {author} {\bibinfo {author} {\bibfnamefont {T.~L.}\ \bibnamefont
  {Hughes}}, \bibinfo {author} {\bibfnamefont {E.}~\bibnamefont {Prodan}},\
  and\ \bibinfo {author} {\bibfnamefont {B.~A.}\ \bibnamefont {Bernevig}},\
  }\bibfield  {title} {\bibinfo {title} {Inversion-symmetric topological
  insulators},\ }\href {https://doi.org/10.1103/PhysRevB.83.245132} {\bibfield
  {journal} {\bibinfo  {journal} {Physical Review B}\ }\textbf {\bibinfo
  {volume} {83}},\ \bibinfo {pages} {245132} (\bibinfo {year}
  {2011})}\BibitemShut {NoStop}%
\bibitem [{\citenamefont {Fu}(2011)}]{tci_3}%
  \BibitemOpen
  \bibfield  {author} {\bibinfo {author} {\bibfnamefont {L.}~\bibnamefont
  {Fu}},\ }\bibfield  {title} {\bibinfo {title} {Topological crystalline
  insulators},\ }\href {https://doi.org/10.1103/physrevlett.106.106802}
  {\bibfield  {journal} {\bibinfo  {journal} {Physical Review Letters}\
  }\textbf {\bibinfo {volume} {106}},\ \bibinfo {pages} {106802} (\bibinfo
  {year} {2011})}\BibitemShut {NoStop}%
\bibitem [{\citenamefont {Slager}\ \emph {et~al.}(2013)\citenamefont {Slager},
  \citenamefont {Mesaros}, \citenamefont {Juri\v{c}i\'c},\ and\ \citenamefont
  {Zaanen}}]{Slager1}%
  \BibitemOpen
  \bibfield  {author} {\bibinfo {author} {\bibfnamefont {R.-J.}\ \bibnamefont
  {Slager}}, \bibinfo {author} {\bibfnamefont {A.}~\bibnamefont {Mesaros}},
  \bibinfo {author} {\bibfnamefont {V.}~\bibnamefont {Juri\v{c}i\'c}},\ and\
  \bibinfo {author} {\bibfnamefont {J.}~\bibnamefont {Zaanen}},\ }\bibfield
  {title} {\bibinfo {title} {The space group classification of topological
  band-insulators},\ }\href {https://doi.org/10.1038/nphys2513} {\bibfield
  {journal} {\bibinfo  {journal} {Nat. Phys.}\ }\textbf {\bibinfo {volume}
  {9}},\ \bibinfo {pages} {98} (\bibinfo {year} {2013})}\BibitemShut {NoStop}%
\bibitem [{\citenamefont {Kruthoff}\ \emph {et~al.}(2017)\citenamefont
  {Kruthoff}, \citenamefont {de~Boer}, \citenamefont {van Wezel}, \citenamefont
  {Kane},\ and\ \citenamefont {Slager}}]{Slager2}%
  \BibitemOpen
  \bibfield  {author} {\bibinfo {author} {\bibfnamefont {J.}~\bibnamefont
  {Kruthoff}}, \bibinfo {author} {\bibfnamefont {J.}~\bibnamefont {de~Boer}},
  \bibinfo {author} {\bibfnamefont {J.}~\bibnamefont {van Wezel}}, \bibinfo
  {author} {\bibfnamefont {C.~L.}\ \bibnamefont {Kane}},\ and\ \bibinfo
  {author} {\bibfnamefont {R.-J.}\ \bibnamefont {Slager}},\ }\bibfield  {title}
  {\bibinfo {title} {Topological classification of crystalline insulators
  through band structure combinatorics},\ }\href
  {https://doi.org/10.1103/PhysRevX.7.041069} {\bibfield  {journal} {\bibinfo
  {journal} {Phys. Rev. X}\ }\textbf {\bibinfo {volume} {7}},\ \bibinfo {pages}
  {041069} (\bibinfo {year} {2017})}\BibitemShut {NoStop}%
\bibitem [{\citenamefont {Benalcazar}\ \emph {et~al.}(2017)\citenamefont
  {Benalcazar}, \citenamefont {Bernevig},\ and\ \citenamefont
  {Hughes}}]{tci_1}%
  \BibitemOpen
  \bibfield  {author} {\bibinfo {author} {\bibfnamefont {W.~A.}\ \bibnamefont
  {Benalcazar}}, \bibinfo {author} {\bibfnamefont {B.~A.}\ \bibnamefont
  {Bernevig}},\ and\ \bibinfo {author} {\bibfnamefont {T.~L.}\ \bibnamefont
  {Hughes}},\ }\bibfield  {title} {\bibinfo {title} {Quantized electric
  multipole insulators},\ }\href {https://doi.org/10.1126/science.aah6442}
  {\bibfield  {journal} {\bibinfo  {journal} {Science}\ }\textbf {\bibinfo
  {volume} {357}},\ \bibinfo {pages} {61} (\bibinfo {year} {2017})}\BibitemShut
  {NoStop}%
\bibitem [{\citenamefont {Hafezi}\ \emph {et~al.}(2013)\citenamefont {Hafezi},
  \citenamefont {Mittal}, \citenamefont {Fan}, \citenamefont {Migdall},\ and\
  \citenamefont {Taylor}}]{ef_1}%
  \BibitemOpen
  \bibfield  {author} {\bibinfo {author} {\bibfnamefont {M.}~\bibnamefont
  {Hafezi}}, \bibinfo {author} {\bibfnamefont {S.}~\bibnamefont {Mittal}},
  \bibinfo {author} {\bibfnamefont {J.}~\bibnamefont {Fan}}, \bibinfo {author}
  {\bibfnamefont {A.}~\bibnamefont {Migdall}},\ and\ \bibinfo {author}
  {\bibfnamefont {J.~M.}\ \bibnamefont {Taylor}},\ }\bibfield  {title}
  {\bibinfo {title} {Imaging topological edge states in silicon photonics},\
  }\href {https://doi.org/10.1038/nphoton.2013.274} {\bibfield  {journal}
  {\bibinfo  {journal} {Nature Photonics}\ }\textbf {\bibinfo {volume} {7}},\
  \bibinfo {pages} {1001} (\bibinfo {year} {2013})}\BibitemShut {NoStop}%
\bibitem [{\citenamefont {Poo}\ \emph {et~al.}(2011)\citenamefont {Poo},
  \citenamefont {Wu}, \citenamefont {Lin}, \citenamefont {Yang},\ and\
  \citenamefont {Chan}}]{ef_2}%
  \BibitemOpen
  \bibfield  {author} {\bibinfo {author} {\bibfnamefont {Y.}~\bibnamefont
  {Poo}}, \bibinfo {author} {\bibfnamefont {R.-X.}\ \bibnamefont {Wu}},
  \bibinfo {author} {\bibfnamefont {Z.}~\bibnamefont {Lin}}, \bibinfo {author}
  {\bibfnamefont {Y.}~\bibnamefont {Yang}},\ and\ \bibinfo {author}
  {\bibfnamefont {C.~T.}\ \bibnamefont {Chan}},\ }\bibfield  {title} {\bibinfo
  {title} {Experimental realization of self-guiding unidirectional
  electromagnetic edge states},\ }\href
  {https://doi.org/10.1103/physrevlett.106.093903} {\bibfield  {journal}
  {\bibinfo  {journal} {Physical Review Letters}\ }\textbf {\bibinfo {volume}
  {106}},\ \bibinfo {pages} {093903} (\bibinfo {year} {2011})}\BibitemShut
  {NoStop}%
\bibitem [{\citenamefont {Wang}\ \emph {et~al.}(2009)\citenamefont {Wang},
  \citenamefont {Chong}, \citenamefont {Joannopoulos},\ and\ \citenamefont
  {Soljačić}}]{ef_3}%
  \BibitemOpen
  \bibfield  {author} {\bibinfo {author} {\bibfnamefont {Z.}~\bibnamefont
  {Wang}}, \bibinfo {author} {\bibfnamefont {Y.}~\bibnamefont {Chong}},
  \bibinfo {author} {\bibfnamefont {J.~D.}\ \bibnamefont {Joannopoulos}},\ and\
  \bibinfo {author} {\bibfnamefont {M.}~\bibnamefont {Soljačić}},\ }\bibfield
   {title} {\bibinfo {title} {Observation of unidirectional
  backscattering-immune topological electromagnetic states},\ }\href
  {https://doi.org/10.1038/nature08293} {\bibfield  {journal} {\bibinfo
  {journal} {Nature}\ }\textbf {\bibinfo {volume} {461}},\ \bibinfo {pages}
  {772} (\bibinfo {year} {2009})}\BibitemShut {NoStop}%
\bibitem [{\citenamefont {Hsieh}\ \emph {et~al.}(2008)\citenamefont {Hsieh},
  \citenamefont {Qian}, \citenamefont {Wray}, \citenamefont {Xia},
  \citenamefont {Hor}, \citenamefont {Cava},\ and\ \citenamefont
  {Hasan}}]{tis_1}%
  \BibitemOpen
  \bibfield  {author} {\bibinfo {author} {\bibfnamefont {D.}~\bibnamefont
  {Hsieh}}, \bibinfo {author} {\bibfnamefont {D.}~\bibnamefont {Qian}},
  \bibinfo {author} {\bibfnamefont {L.}~\bibnamefont {Wray}}, \bibinfo {author}
  {\bibfnamefont {Y.}~\bibnamefont {Xia}}, \bibinfo {author} {\bibfnamefont
  {Y.~S.}\ \bibnamefont {Hor}}, \bibinfo {author} {\bibfnamefont {R.~J.}\
  \bibnamefont {Cava}},\ and\ \bibinfo {author} {\bibfnamefont {M.~Z.}\
  \bibnamefont {Hasan}},\ }\bibfield  {title} {\bibinfo {title} {A topological
  dirac insulator in a quantum spin hall phase},\ }\href
  {https://doi.org/10.1038/nature06843} {\bibfield  {journal} {\bibinfo
  {journal} {Nature}\ }\textbf {\bibinfo {volume} {452}},\ \bibinfo {pages}
  {970} (\bibinfo {year} {2008})}\BibitemShut {NoStop}%
\bibitem [{\citenamefont {Konig}\ \emph {et~al.}(2007)\citenamefont {Konig},
  \citenamefont {Wiedmann}, \citenamefont {Brune}, \citenamefont {Roth},
  \citenamefont {Buhmann}, \citenamefont {Molenkamp}, \citenamefont {Qi},\ and\
  \citenamefont {Zhang}}]{tis_2}%
  \BibitemOpen
  \bibfield  {author} {\bibinfo {author} {\bibfnamefont {M.}~\bibnamefont
  {Konig}}, \bibinfo {author} {\bibfnamefont {S.}~\bibnamefont {Wiedmann}},
  \bibinfo {author} {\bibfnamefont {C.}~\bibnamefont {Brune}}, \bibinfo
  {author} {\bibfnamefont {A.}~\bibnamefont {Roth}}, \bibinfo {author}
  {\bibfnamefont {H.}~\bibnamefont {Buhmann}}, \bibinfo {author} {\bibfnamefont
  {L.~W.}\ \bibnamefont {Molenkamp}}, \bibinfo {author} {\bibfnamefont {X.-L.}\
  \bibnamefont {Qi}},\ and\ \bibinfo {author} {\bibfnamefont {S.-C.}\
  \bibnamefont {Zhang}},\ }\bibfield  {title} {\bibinfo {title} {Quantum spin
  hall insulator state in hgte quantum wells},\ }\href
  {https://doi.org/10.1126/science.1148047} {\bibfield  {journal} {\bibinfo
  {journal} {Science}\ }\textbf {\bibinfo {volume} {318}},\ \bibinfo {pages}
  {766} (\bibinfo {year} {2007})}\BibitemShut {NoStop}%
\bibitem [{\citenamefont {Bernevig}\ \emph {et~al.}(2006)\citenamefont
  {Bernevig}, \citenamefont {Hughes},\ and\ \citenamefont {Zhang}}]{tis_3}%
  \BibitemOpen
  \bibfield  {author} {\bibinfo {author} {\bibfnamefont {B.~A.}\ \bibnamefont
  {Bernevig}}, \bibinfo {author} {\bibfnamefont {T.~L.}\ \bibnamefont
  {Hughes}},\ and\ \bibinfo {author} {\bibfnamefont {S.-C.}\ \bibnamefont
  {Zhang}},\ }\bibfield  {title} {\bibinfo {title} {Quantum spin hall effect
  and topological phase transition in hgte quantum wells},\ }\href
  {https://doi.org/10.1126/science.1133734} {\bibfield  {journal} {\bibinfo
  {journal} {Science}\ }\textbf {\bibinfo {volume} {314}},\ \bibinfo {pages}
  {1757} (\bibinfo {year} {2006})}\BibitemShut {NoStop}%
\bibitem [{\citenamefont {Kane}\ and\ \citenamefont
  {Mele}(2005{\natexlab{a}})}]{tis_4}%
  \BibitemOpen
  \bibfield  {author} {\bibinfo {author} {\bibfnamefont {C.~L.}\ \bibnamefont
  {Kane}}\ and\ \bibinfo {author} {\bibfnamefont {E.~J.}\ \bibnamefont
  {Mele}},\ }\bibfield  {title} {\bibinfo {title} {Z2topological order and the
  quantum spin hall effect},\ }\href
  {https://doi.org/10.1103/physrevlett.95.146802} {\bibfield  {journal}
  {\bibinfo  {journal} {Physical Review Letters}\ }\textbf {\bibinfo {volume}
  {95}},\ \bibinfo {pages} {146802} (\bibinfo {year}
  {2005}{\natexlab{a}})}\BibitemShut {NoStop}%
\bibitem [{\citenamefont {Kane}\ and\ \citenamefont
  {Mele}(2005{\natexlab{b}})}]{tis_5}%
  \BibitemOpen
  \bibfield  {author} {\bibinfo {author} {\bibfnamefont {C.~L.}\ \bibnamefont
  {Kane}}\ and\ \bibinfo {author} {\bibfnamefont {E.~J.}\ \bibnamefont
  {Mele}},\ }\bibfield  {title} {\bibinfo {title} {Quantum spin hall effect in
  graphene},\ }\href {https://doi.org/10.1103/physrevlett.95.226801} {\bibfield
   {journal} {\bibinfo  {journal} {Physical Review Letters}\ }\textbf {\bibinfo
  {volume} {95}},\ \bibinfo {pages} {226801} (\bibinfo {year}
  {2005}{\natexlab{b}})}\BibitemShut {NoStop}%
\bibitem [{\citenamefont {Laughlin}(1983)}]{tis_6}%
  \BibitemOpen
  \bibfield  {author} {\bibinfo {author} {\bibfnamefont {R.~B.}\ \bibnamefont
  {Laughlin}},\ }\bibfield  {title} {\bibinfo {title} {Anomalous quantum hall
  effect: An incompressible quantum fluid with fractionally charged
  excitations},\ }\href {https://doi.org/10.1103/physrevlett.50.1395}
  {\bibfield  {journal} {\bibinfo  {journal} {Physical Review Letters}\
  }\textbf {\bibinfo {volume} {50}},\ \bibinfo {pages} {1395} (\bibinfo {year}
  {1983})}\BibitemShut {NoStop}%
\bibitem [{\citenamefont {Klitzing}\ \emph {et~al.}(1980)\citenamefont
  {Klitzing}, \citenamefont {Dorda},\ and\ \citenamefont {Pepper}}]{tis_7}%
  \BibitemOpen
  \bibfield  {author} {\bibinfo {author} {\bibfnamefont {K.~V.}\ \bibnamefont
  {Klitzing}}, \bibinfo {author} {\bibfnamefont {G.}~\bibnamefont {Dorda}},\
  and\ \bibinfo {author} {\bibfnamefont {M.}~\bibnamefont {Pepper}},\
  }\bibfield  {title} {\bibinfo {title} {New method for high-accuracy
  determination of the fine-structure constant based on quantized hall
  resistance},\ }\href {https://doi.org/10.1103/physrevlett.45.494} {\bibfield
  {journal} {\bibinfo  {journal} {Physical Review Letters}\ }\textbf {\bibinfo
  {volume} {45}},\ \bibinfo {pages} {494} (\bibinfo {year} {1980})}\BibitemShut
  {NoStop}%
\bibitem [{\citenamefont {Yang}\ \emph
  {et~al.}(2020{\natexlab{a}})\citenamefont {Yang}, \citenamefont {Li},
  \citenamefont {Peng}, \citenamefont {Zou},\ and\ \citenamefont
  {Cheng}}]{YZZ}%
  \BibitemOpen
  \bibfield  {author} {\bibinfo {author} {\bibfnamefont {Z.-Z.}\ \bibnamefont
  {Yang}}, \bibinfo {author} {\bibfnamefont {X.}~\bibnamefont {Li}}, \bibinfo
  {author} {\bibfnamefont {Y.-Y.}\ \bibnamefont {Peng}}, \bibinfo {author}
  {\bibfnamefont {X.-Y.}\ \bibnamefont {Zou}},\ and\ \bibinfo {author}
  {\bibfnamefont {J.-C.}\ \bibnamefont {Cheng}},\ }\bibfield  {title} {\bibinfo
  {title} {Helical higher-order topological states in an acoustic crystalline
  insulator},\ }\href {https://doi.org/10.1103/PhysRevLett.125.255502}
  {\bibfield  {journal} {\bibinfo  {journal} {Phys. Rev. Lett.}\ }\textbf
  {\bibinfo {volume} {125}},\ \bibinfo {pages} {255502} (\bibinfo {year}
  {2020}{\natexlab{a}})}\BibitemShut {NoStop}%
\bibitem [{\citenamefont {Freeney}\ \emph {et~al.}(2020)\citenamefont
  {Freeney}, \citenamefont {Van Den~Broeke}, \citenamefont {Harsveld Van
  Der~Veen}, \citenamefont {Swart},\ and\ \citenamefont {Morais~Smith}}]{ba_1}%
  \BibitemOpen
  \bibfield  {author} {\bibinfo {author} {\bibfnamefont {S.}~\bibnamefont
  {Freeney}}, \bibinfo {author} {\bibfnamefont {J.}~\bibnamefont {Van
  Den~Broeke}}, \bibinfo {author} {\bibfnamefont {A.}~\bibnamefont {Harsveld
  Van Der~Veen}}, \bibinfo {author} {\bibfnamefont {I.}~\bibnamefont {Swart}},\
  and\ \bibinfo {author} {\bibfnamefont {C.}~\bibnamefont {Morais~Smith}},\
  }\bibfield  {title} {\bibinfo {title} {Edge-dependent topology in kekulé
  lattices},\ }\href {https://doi.org/10.1103/physrevlett.124.236404}
  {\bibfield  {journal} {\bibinfo  {journal} {Physical Review Letters}\
  }\textbf {\bibinfo {volume} {124}},\ \bibinfo {pages} {236404} (\bibinfo
  {year} {2020})}\BibitemShut {NoStop}%
\bibitem [{\citenamefont {Chen}\ \emph
  {et~al.}(2019{\natexlab{b}})\citenamefont {Chen}, \citenamefont {Xu},
  \citenamefont {Al~Jahdali}, \citenamefont {Mei},\ and\ \citenamefont
  {Wu}}]{ba_2}%
  \BibitemOpen
  \bibfield  {author} {\bibinfo {author} {\bibfnamefont {Z.-G.}\ \bibnamefont
  {Chen}}, \bibinfo {author} {\bibfnamefont {C.}~\bibnamefont {Xu}}, \bibinfo
  {author} {\bibfnamefont {R.}~\bibnamefont {Al~Jahdali}}, \bibinfo {author}
  {\bibfnamefont {J.}~\bibnamefont {Mei}},\ and\ \bibinfo {author}
  {\bibfnamefont {Y.}~\bibnamefont {Wu}},\ }\bibfield  {title} {\bibinfo
  {title} {Corner states in a second-order acoustic topological insulator as
  bound states in the continuum},\ }\href
  {https://doi.org/10.1103/physrevb.100.075120} {\bibfield  {journal} {\bibinfo
   {journal} {Physical Review B}\ }\textbf {\bibinfo {volume} {100}},\ \bibinfo
  {pages} {075120} (\bibinfo {year} {2019}{\natexlab{b}})}\BibitemShut
  {NoStop}%
\bibitem [{\citenamefont {Song}\ \emph {et~al.}(2017)\citenamefont {Song},
  \citenamefont {Fang},\ and\ \citenamefont {Fang}}]{ba_3}%
  \BibitemOpen
  \bibfield  {author} {\bibinfo {author} {\bibfnamefont {Z.}~\bibnamefont
  {Song}}, \bibinfo {author} {\bibfnamefont {Z.}~\bibnamefont {Fang}},\ and\
  \bibinfo {author} {\bibfnamefont {C.}~\bibnamefont {Fang}},\ }\bibfield
  {title} {\bibinfo {title} {(d-2)-dimensional edge states of rotation symmetry
  protected topological states},\ }\href
  {https://doi.org/10.1103/physrevlett.119.246402} {\bibfield  {journal}
  {\bibinfo  {journal} {Physical Review Letters}\ }\textbf {\bibinfo {volume}
  {119}},\ \bibinfo {pages} {246402} (\bibinfo {year} {2017})}\BibitemShut
  {NoStop}%
\bibitem [{\citenamefont {Langbehn}\ \emph {et~al.}(2017)\citenamefont
  {Langbehn}, \citenamefont {Peng}, \citenamefont {Trifunovic}, \citenamefont
  {Von~Oppen},\ and\ \citenamefont {Brouwer}}]{ba_4}%
  \BibitemOpen
  \bibfield  {author} {\bibinfo {author} {\bibfnamefont {J.}~\bibnamefont
  {Langbehn}}, \bibinfo {author} {\bibfnamefont {Y.}~\bibnamefont {Peng}},
  \bibinfo {author} {\bibfnamefont {L.}~\bibnamefont {Trifunovic}}, \bibinfo
  {author} {\bibfnamefont {F.}~\bibnamefont {Von~Oppen}},\ and\ \bibinfo
  {author} {\bibfnamefont {P.~W.}\ \bibnamefont {Brouwer}},\ }\bibfield
  {title} {\bibinfo {title} {Reflection-symmetric second-order topological
  insulators and superconductors},\ }\href
  {https://doi.org/10.1103/physrevlett.119.246401} {\bibfield  {journal}
  {\bibinfo  {journal} {Physical Review Letters}\ }\textbf {\bibinfo {volume}
  {119}},\ \bibinfo {pages} {246401} (\bibinfo {year} {2017})}\BibitemShut
  {NoStop}%
\bibitem [{\citenamefont {Yang}\ \emph {et~al.}(2021)\citenamefont {Yang},
  \citenamefont {Guan}, \citenamefont {Zou},\ and\ \citenamefont
  {Cheng}}]{add1}%
  \BibitemOpen
  \bibfield  {author} {\bibinfo {author} {\bibfnamefont {Z.-Z.}\ \bibnamefont
  {Yang}}, \bibinfo {author} {\bibfnamefont {A.-Y.}\ \bibnamefont {Guan}},
  \bibinfo {author} {\bibfnamefont {X.-Y.}\ \bibnamefont {Zou}},\ and\ \bibinfo
  {author} {\bibfnamefont {J.-C.}\ \bibnamefont {Cheng}},\ }\href@noop {}
  {\bibinfo {title} {Topological classical systems with generalized chiral
  symmetry}} (\bibinfo {year} {2021}),\ \Eprint
  {https://arxiv.org/abs/2102.10361} {arXiv:2102.10361} \BibitemShut {NoStop}%
\bibitem [{\citenamefont {Li}\ \emph {et~al.}(2019)\citenamefont {Li},
  \citenamefont {Zhirihin}, \citenamefont {Gorlach}, \citenamefont {Ni},
  \citenamefont {Filonov}, \citenamefont {Slobozhanyuk}, \citenamefont {Alù},\
  and\ \citenamefont {Khanikaev}}]{tpk_1}%
  \BibitemOpen
  \bibfield  {author} {\bibinfo {author} {\bibfnamefont {M.}~\bibnamefont
  {Li}}, \bibinfo {author} {\bibfnamefont {D.}~\bibnamefont {Zhirihin}},
  \bibinfo {author} {\bibfnamefont {M.}~\bibnamefont {Gorlach}}, \bibinfo
  {author} {\bibfnamefont {X.}~\bibnamefont {Ni}}, \bibinfo {author}
  {\bibfnamefont {D.}~\bibnamefont {Filonov}}, \bibinfo {author} {\bibfnamefont
  {A.}~\bibnamefont {Slobozhanyuk}}, \bibinfo {author} {\bibfnamefont
  {A.}~\bibnamefont {Alù}},\ and\ \bibinfo {author} {\bibfnamefont {A.~B.}\
  \bibnamefont {Khanikaev}},\ }\bibfield  {title} {\bibinfo {title}
  {Higher-order topological states in photonic kagome crystals with long-range
  interactions},\ }\href {https://doi.org/10.1038/s41566-019-0561-9} {\bibfield
   {journal} {\bibinfo  {journal} {Nature Photonics}\ }\textbf {\bibinfo
  {volume} {14}},\ \bibinfo {pages} {89} (\bibinfo {year} {2019})}\BibitemShut
  {NoStop}%
\bibitem [{\citenamefont {Chen}\ \emph
  {et~al.}(2019{\natexlab{c}})\citenamefont {Chen}, \citenamefont {Xu},
  \citenamefont {Al~Jahdali}, \citenamefont {Mei},\ and\ \citenamefont
  {Wu}}]{tpk_2}%
  \BibitemOpen
  \bibfield  {author} {\bibinfo {author} {\bibfnamefont {Z.-G.}\ \bibnamefont
  {Chen}}, \bibinfo {author} {\bibfnamefont {C.}~\bibnamefont {Xu}}, \bibinfo
  {author} {\bibfnamefont {R.}~\bibnamefont {Al~Jahdali}}, \bibinfo {author}
  {\bibfnamefont {J.}~\bibnamefont {Mei}},\ and\ \bibinfo {author}
  {\bibfnamefont {Y.}~\bibnamefont {Wu}},\ }\bibfield  {title} {\bibinfo
  {title} {Corner states in a second-order acoustic topological insulator as
  bound states in the continuum},\ }\href
  {https://doi.org/10.1103/PhysRevB.100.075120} {\bibfield  {journal} {\bibinfo
   {journal} {Physical Review B}\ }\textbf {\bibinfo {volume} {100}},\ \bibinfo
  {pages} {075120} (\bibinfo {year} {2019}{\natexlab{c}})}\BibitemShut
  {NoStop}%
\bibitem [{\citenamefont {Liu}\ and\ \citenamefont
  {Wakabayashi}(2017)}]{tpk_3}%
  \BibitemOpen
  \bibfield  {author} {\bibinfo {author} {\bibfnamefont {F.}~\bibnamefont
  {Liu}}\ and\ \bibinfo {author} {\bibfnamefont {K.}~\bibnamefont
  {Wakabayashi}},\ }\bibfield  {title} {\bibinfo {title} {Novel topological
  phase with a zero berry curvature},\ }\href
  {https://doi.org/10.1103/PhysRevLett.118.076803} {\bibfield  {journal}
  {\bibinfo  {journal} {Physical Review Letters}\ }\textbf {\bibinfo {volume}
  {118}},\ \bibinfo {pages} {076803} (\bibinfo {year} {2017})}\BibitemShut
  {NoStop}%
\bibitem [{\citenamefont {Benalcazar}\ \emph {et~al.}(2019)\citenamefont
  {Benalcazar}, \citenamefont {Li},\ and\ \citenamefont {Hughes}}]{cs_1}%
  \BibitemOpen
  \bibfield  {author} {\bibinfo {author} {\bibfnamefont {W.~A.}\ \bibnamefont
  {Benalcazar}}, \bibinfo {author} {\bibfnamefont {T.}~\bibnamefont {Li}},\
  and\ \bibinfo {author} {\bibfnamefont {T.~L.}\ \bibnamefont {Hughes}},\
  }\bibfield  {title} {\bibinfo {title} {Quantization of fractional corner
  charge in $c_n$-symmetric higher-order topological crystalline insulators},\
  }\href {https://doi.org/10.1103/PhysRevB.99.245151} {\bibfield  {journal}
  {\bibinfo  {journal} {Physical Review B}\ }\textbf {\bibinfo {volume} {99}},\
  \bibinfo {pages} {245151} (\bibinfo {year} {2019})}\BibitemShut {NoStop}%
\bibitem [{\citenamefont {Beranek}(1954)}]{analogy}%
  \BibitemOpen
  \bibfield  {author} {\bibinfo {author} {\bibfnamefont {L.~L.}\ \bibnamefont
  {Beranek}},\ }\href@noop {} {\emph {\bibinfo {title} {Acoustics}}},\ New
  York, McGraw-Hill\ (\bibinfo {address} {New York,},\ \bibinfo {year}
  {1954})\BibitemShut {NoStop}%
\bibitem [{\citenamefont {Yang}\ \emph
  {et~al.}(2020{\natexlab{b}})\citenamefont {Yang}, \citenamefont {Peng},
  \citenamefont {Li}, \citenamefont {Zou},\ and\ \citenamefont
  {Cheng}}]{ribbon}%
  \BibitemOpen
  \bibfield  {author} {\bibinfo {author} {\bibfnamefont {Z.-Z.}\ \bibnamefont
  {Yang}}, \bibinfo {author} {\bibfnamefont {Y.-Y.}\ \bibnamefont {Peng}},
  \bibinfo {author} {\bibfnamefont {X.}~\bibnamefont {Li}}, \bibinfo {author}
  {\bibfnamefont {X.-Y.}\ \bibnamefont {Zou}},\ and\ \bibinfo {author}
  {\bibfnamefont {J.-C.}\ \bibnamefont {Cheng}},\ }\bibfield  {title} {\bibinfo
  {title} {Boundary-dependent corner states in topological acoustic resonator
  array},\ }\href {https://doi.org/10.1063/5.0017503} {\bibfield  {journal}
  {\bibinfo  {journal} {Applied Physics Letters}\ }\textbf {\bibinfo {volume}
  {117}},\ \bibinfo {pages} {113501} (\bibinfo {year}
  {2020}{\natexlab{b}})}\BibitemShut {NoStop}%
\bibitem [{\citenamefont {Peterson}\ \emph {et~al.}(2020)\citenamefont
  {Peterson}, \citenamefont {Li}, \citenamefont {Benalcazar}, \citenamefont
  {Hughes},\ and\ \citenamefont {Bahl}}]{add2}%
  \BibitemOpen
  \bibfield  {author} {\bibinfo {author} {\bibfnamefont {C.~W.}\ \bibnamefont
  {Peterson}}, \bibinfo {author} {\bibfnamefont {T.}~\bibnamefont {Li}},
  \bibinfo {author} {\bibfnamefont {W.~A.}\ \bibnamefont {Benalcazar}},
  \bibinfo {author} {\bibfnamefont {T.~L.}\ \bibnamefont {Hughes}},\ and\
  \bibinfo {author} {\bibfnamefont {G.}~\bibnamefont {Bahl}},\ }\bibfield
  {title} {\bibinfo {title} {A fractional corner anomaly reveals higher-order
  topology},\ }\href {https://science.sciencemag.org/content/368/6495/1114}
  {\bibfield  {journal} {\bibinfo  {journal} {Science}\ }\textbf {\bibinfo
  {volume} {368}},\ \bibinfo {pages} {1114} (\bibinfo {year}
  {2020})}\BibitemShut {NoStop}%
\end{thebibliography}
\providecommand{\noopsort}[1]{}\providecommand{\singleletter}[1]{#1}%

\newpage

\end{document}